\documentclass[twocolumn]{aastex63}

\usepackage{bm}

\shorttitle{}
\shortauthors{Kimura, Takasao, Tomida}

\begin{document}
\title{Modeling Hadronic Gamma-ray Emissions from Solar Flares \\
and Prospects for Detecting Non-thermal Signatures from Protostars
}

\correspondingauthor{Shigeo S. Kimura}
\email{shigeo@astr.tohoku.ac.jp}

\author[0000-0003-2579-7266]{Shigeo S. Kimura}
\affiliation{Frontier Research Institute for Interdisciplinary Sciences, Tohoku University, Sendai 980-8578, Japan}
\affiliation{Astronomical Institute, Graduate School of Science, Tohoku University, Sendai 980-8578, Japan}

\author[0000-0003-3882-3945]{Shinsuke Takasao}
\affil{Department of Earth and Space Science, Osaka University, Toyonaka, Osaka, 560-0043, Japan}
\author[0000-0001-8105-8113]{Kengo Tomida}
\affiliation{Astronomical Institute, Graduate School of Science, Tohoku University, Sendai 980-8578, Japan}




\begin{abstract}
We investigate gamma-ray emission in the impulsive phase of solar flares and the detectability of non-thermal signatures from protostellar flares. Energetic solar flares emit high-energy gamma rays of GeV energies, but their production mechanism and emission site are still unknown. Young stellar objects, including protostars, also exhibit luminous X-ray flares, but the triggering mechanism of the flaring activity is still unclear due to the strong obscuration. Non-thermal signatures in mm/sub-mm and gamma-ray bands are useful to probe protostellar flares owing to their strong penetration power.
We develop a non-thermal emission model of the impulsive phase of solar flares, where cosmic-ray protons accelerated at the termination shock produce high-energy gamma rays via hadronuclear interaction with the evaporation plasma. This model can reproduce gamma-ray data in the impulsive phase of a solar flare.
We apply our model to protostellar flares and show that Cherenkov Telescope Array will be able to detect gamma rays of TeV energies if particle acceleration in protostellar flares is efficient. Non-thermal electrons accelerated together with protons can emit strong mm and sub-mm signals via synchrotron radiation, whose power is consistent with the energetic mm/sub-mm transients observed from young stars. Future gamma-ray and mm/sub-mm observations from protostars, coordinated with a hard X-ray observation, will unravel the  non-thermal particle production and triggering mechanism of protostellar flares.
\end{abstract}



%
\section{Introduction}

Solar flares are transient phenomena caused by sudden magnetic energy release at the solar surface \citep[see e.g.,][for reviews]{2011LRSP....8....6S,2011SSRv..158....5H,2014LRSP...11....3C,2017LRSP...14....2B}. They produce broadband photons from radio to high-energy ($>100$ MeV) gamma rays by various emission mechanisms. The magnetic reconnection heats up the plasma in the corona and chromosphere, which results in bright optical/ultraviolet/soft X-ray emission. The magnetic reconnection also accelerates non-thermal electrons, which emit hard X-ray and radio signals \citep{1994Natur.371..495M,2015Sci...350.1238C,2018SSRv..214...82O}.

The Fermi satellite has been detecting high-energy gamma rays  ($E_\gamma>100$ MeV) from solar flares \citep{2021ApJS..252...13A}. The gamma-ray emission episodes can be divided into two phases: the impulsive and gradual phases lasting for $\sim3-10$ minutes and $\sim2-20$ hours, respectively.
In both phases, the high-energy gamma-ray flux is higher than the extrapolation of the hard X-ray power-law spectrum \citep[e.g.,][]{2017ApJ...835..219A},
which indicates that the production mechanism of the gamma rays differs from that of the hard X-rays. Hadronic emissions by energetic protons is a viable mechanism of the gamma-ray production \citep[e.g.,][]{2018ApJ...864..148K}, suggesting that the solar flares can accelerate protons up to GeV energies. However, the proton acceleration and subsequent gamma-ray emission mechanisms are still unclear.
The long duration of the gradual phase might favor continuous proton acceleration, which can be related to coronal mass ejection (CME) and its subsequent interaction with the ambient plasma \citep[e.g.,][]{2014ApJ...789...20A,2016ApJ...830...28P}. On the other hand, recent multi-wavelength analysis suggests multiple particle acceleration sites during the solar flares, and the termination shock at the loop top is a plausible proton acceleration site in the impulsive phase \citep{2021ApJ...915...12K}. The non-thermal electron production at the termination shock is actively discussed \citep{1994Natur.371..495M,2015Sci...350.1238C}, but the proton acceleration and gamma-ray emission there has yet to be investigated in detail.

Young stellar objects (YSOs) also exhibit flaring activities in X-ray bands in all the stages of their evolution: from the protostar phase (class-0/class-I) to the pre-main sequence phase (class-II/class-III) \citep{1996PASJ...48L..87K,1999ARA&A..37..363F,2002ApJ...574..258F,2007A&A...468..463S,2020A&A...638L...4G}. Compared to solar flares, X-ray flares from YSOs are energetic and frequent. Their total X-ray energy reaches $\mathcal{E}_X\simeq10^{35}-10^{37}\rm~erg$,
which is likely a small fraction of the total radiation energy and released magnetic energy \citep[see, e.g.,][for solar flares]{2012ApJ...759...71E}. This indicates that protostellar flares are about $10^4-10^6$ times more energetic than the largest recorded solar flares \citep[e.g.,][]{2013JSWSC...3A..31C}.
However, the triggering mechanism of the YSO flares, especially in the protostellar phase, are still controversial. 

YSO flares affect the thermochemical structure in the star forming regions by producing strong X-rays. Theoretical studies argue that strong X-rays can change the abundance ratio of molecules such as gas-phase H$_2$O in the protostellar envelopes \citep{2005A&A...440..949S,2006A&A...453..555S,2021A&A...650A.180N} and protoplanetary disks \citep{2019ApJ...883..197W}. X-rays by YSO flares can affect the location of snow lines, and recent radio observations suggesting the lack of water vapor can be consistent with the argument \citep{2020A&A...636A..26H}.

YSO flares are also expected to accelerate cosmic rays (CRs), which also affect the chemical structure by ionizing the gas and destructing molecules \citep[see, e.g.,][for a recent review]{2020SSRv..216...29P}. Nevertheless, the CR production efficiency in the YSO flares are unknown. Since YSO flares are much more energetic than solar flares, we expect that CR particles can be accelerated to much higher energies. CR protons emit gamma rays via hadronuclear interactions, whereas CR electrons emit mm/sub-mm signals via synchrotron radiation. These signals do not suffer from attenuation even in the dense cold medium, and thus, they are powerful tools to probe the flaring activities and CR productions. Theoretical modelings of non-thermal signatures from YSOs, including emission from jets and magnetospheres, have been discussed \citep[e.g.,][]{2007A&A...476.1289A,2011ApJ...738..115D,2021MNRAS.504.2405A}, but they focus on a specific T-Tauri star or emission from protostellar jets. Detectability of non-thermal signatures from YSO flares in general are not investigated in detail so far.

In this paper, we construct a model of hadronic gamma-ray emission from solar flares, and discuss detectability of non-thermal signatures from protostellar flares. 
We first model hadronic gamma-ray emission from the impulsive phase of solar flares and calibrate the model parameters using the solar flare data. Then, we apply our model to protostellar flares, and discuss detectability of radio and gamma-ray signals by current and future detectors. Comparison of model prediction with future radio and gamma-ray data will unravel the CR production mechanism in protostellar flares, which may help understand the effects of CRs during the star and planet formation. 
This paper is organized as follows. In Section \ref{sec:sun}, we describe our gamma-ray emission model of solar flares. We apply the model to protostellar flares in Section \ref{sec:protostar}.
We estimate the synchrotron emission by non-thermal electrons accelerated together with protons in Section \ref{sec:leptonic}. Section \ref{sec:discussion} discusses implications of our results, and Section \ref{sec:summary} summarizes our results. We use convention of $A_X=A/10^X$ in cgs unit unless otherwise noted.

\section{Impulsive Gamma-ray Emission Model for Solar Flares}\label{sec:sun}

  \begin{figure*}
      \centering
       \includegraphics[width=\linewidth]{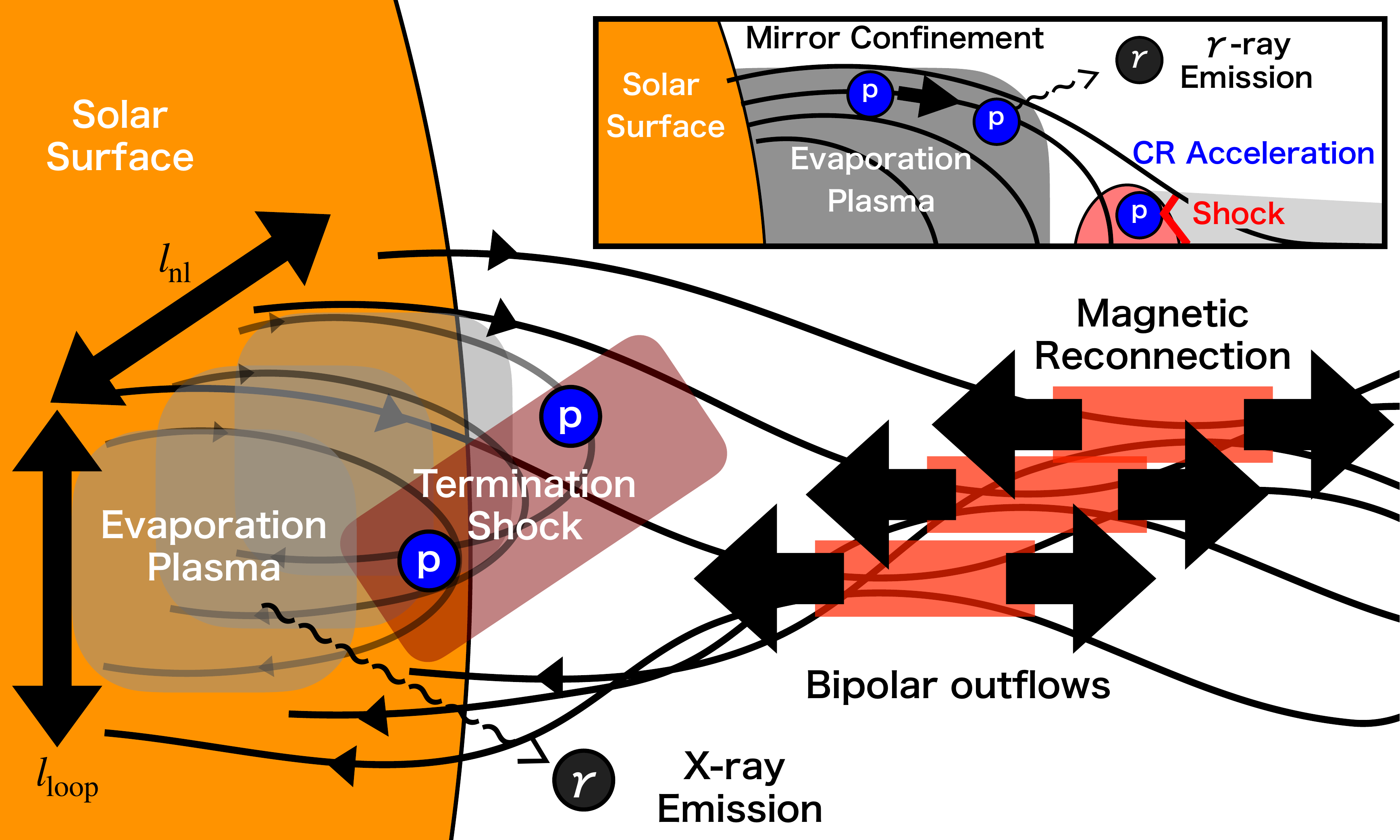}
  \caption{Schematic picture of our hadronic gamma-ray emission model from the impulsive phase of solar flares. The magnetic reconnection event produces bipolar outflows. The anti-sunward outflow may become a CME. The sunward outflow collides with the magnetic field loops,
  which leads to the formation of a termination shock. Non-thermal electron beam and thermal conduction by the shocked gas heat up the chromospheric plasma. This leads to the hot plasma upflows (so-called chromospheric evaporation), and the flare loops are filled with the evaporation plasma. In our model, CR protons are accelerated at the shock and confined inside the flare loop owing to the magnetic mirror effect, as seen in the inset of the figure. These CR protons produce hadronic gamma rays via hadronuclear interactions with the evaporation plasma. }
      \label{fig:schematic_sun}
  \end{figure*}

High-energy gamma rays are accompanied with M-class ($L_X\sim3\times10^{25}-3\times10^{26}\rm~erg~s^{-1}$) and X-class ($L_X\gtrsim3\times10^{26}\rm~erg~s^{-1}$) flares, and the majority of the gamma-ray detected flares indicate high-velocity CME of $\gtrsim10^3\rm~km~s^{-1}$. The peak luminosity of gamma rays from solar flares ranges $L_\gamma\sim5\times10^{18}-5\times10^{21}\rm~erg~s^{-1}$ \citep{2021ApJS..252...13A}. 
These observations suggest a small gamma-ray to X-ray luminosity ratio, $L_\gamma/L_X\sim10^{-9}-10^{-5}$.

The solar flares are divided into two phases \citep{1974IAUS...57..105K}. One is the impulsive phase in which we observe a rapid and strong variability. Prominent features in this phase are strong hard X-ray and radio signals with a typical duration of a few to ten minutes, which are produced by  non-thermal electrons. 
The other is the gradual phase where the lightcurve slowly evolves for a longer timescale. Bright soft X-ray and $\rm H\alpha$ emissions by thermal hot plasma are observed, which lasts a few to several hours.
High-energy gamma rays are accompanied in both phases, though some fraction of gamma-ray flares appear only in either of the phases.

In this section, we construct a model for hadronic gamma-ray emission in the impulsive phase, whose schematic picture is shown in Figure \ref{fig:schematic_sun}.
The magnetic reconnection produces bipolar outflows, and we focus on the sunward outflow colliding with the magnetic field loops. 
We expect the formation of a termination shock via the collision between the reconnection outflow and flare loops, as some observations suggest \citep[e.g.,][]{2002A&A...384..273A,2004ApJ...615..526A}.

If non-thermal electrons are accelerated at the termination shock \citep[e.g.,][]{1998ApJ...495L..67T}, it is natural to consider that the termination shock also accelerates non-thermal protons. These protons interact with dense plasma evaporated from the chromosphere, which results in hadronic high-energy gamma-ray emission.
Throughout this paper, we approximate the flare loop to be a fixed size and filled with uniform plasma for simplicity. In reality, many physical quantities, such as the size of the flare loop and the density of the evaporation plasma, evolve with time. Nevertheless, our treatment provides reasonable agreement with the observational data as shown below.

\subsection{Plasma structure during solar flares}

Observations of X-class flares, defined by the luminosity in the GOES band (1.55--12.4 keV;  $L_X\sim3\times10^{26}-3\times10^{27}\rm~erg~s^{-1}$), revealed that the size of the flare loop and the temperature of evaporation plasma are typically $l_{\rm loop}\sim10^9-10^{10}$ cm and $T_{\rm flare}\sim1\times10^7-4\times10^7$ K, respectively. The durations of the impulsive and gradual phases are typically $t_{\rm imp}\sim10^2-10^3$ sec and $t_{\rm grad}\sim10^3-10^4$ sec, respectively \citep[e.g.,][]{2017LRSP...14....2B}. The magnetic field strength and number density of the coronal region can be $B_{\rm rec}\sim30-300$ G and $n_{\rm rec}\sim10^8-10^9\rm~cm^{-3}$ \citep[e.g.,][]{2011LRSP....8....6S}.  
We search for the appropriate values of $B_{\rm rec}$ and $n_{\rm rec}$ so that the resulting quantities are in agreement with the observations.

Magnetic reconnection produces bipolar outflows whose velocity is roughly equal to the Alfven velocity around the reconnection region (so-called the inflow region):
\begin{equation}
V_{\rm out}\approx V_A=\frac{B_{\rm rec}}{\sqrt{4\pi m_p n_{\rm rec}}}\simeq6.9\times10^8B_{\rm rec,2}n_{\rm rec,9}^{-1/2}\rm~cm~s^{-1},
\end{equation}
where $m_p$ is the proton mass.
The width of the sunward reconnection outflow is estimated to be $l_{\rm out}\sim\eta_{\rm rec}l_{\rm loop}\sim1.0\times10^8l_{\rm loop,9.5}\eta_{\rm rec,-1.5}$ cm, where $\eta_{\rm rec}=V_{\rm in}/V_A$ is the reconnection rate and $V_{\rm in}$ is the reconnection velocity. In solar flares, $\eta_{\rm rec}\sim0.01-0.1$ is observationally estimated \citep[e.g.,][]{2005ApJ...632.1184I,2006ApJ...637.1122N,2012ApJ...745L...6T}. The duration of the reconnection event is given by 
\begin{equation}
 t_{\rm rec}=\frac{l_{\rm loop}}{V_{\rm in}}\approx1.4\times10^2 l_{\rm loop,9.5}n_{\rm rec,9}^{1/2}B_{\rm rec,2}^{-1}\eta_{\rm rec,-1.5}^{-1}\rm~s.
\end{equation}
This is the typical timescale on which the plasma around the reconnection region brings the magnetic field to it. This timescale corresponds to the duration of the impulsive phase, i.e., $t_{\rm imp}\approx t_{\rm rec}$.

The reconnection heats up the outflow plasma, but the heated plasma cools down via the thermal conduction\footnote{Here, we consider the conduction cooling only by thermal electrons for simplicity. In reality,  non-thermal electron beam also carries energy away from the outflow plasma. }. Then, by considering the balance between the reconnection heating and the conduction cooling, the temperature in the outflow is estimated to be  \citep{2001ApJ...549.1160Y,2002ApJ...577..422S}
\begin{equation}
T_{\rm out}\approx\left(\frac{B_{\rm rec}^3l_{\rm loop}}{2 \kappa_0 \sqrt{4\pi m_pn_{\rm rec}}}\right)^{2/7},\label{eq:Tout}
\end{equation}
where we use the Spitzer thermal conduction coefficient, $\kappa=\kappa_0 T^{5/2}$ with $\kappa_0\simeq10^{-6}$ in cgs unit.
With this temperature, the outflow is supersonic, and thus, the outflow forms a termination shock when it collides with the reconnected field lines as long as the guide field is weaker than the reconnecting field. 
The acoustic Mach number of the termination shock is estimated to be \citep{2009ApJ...701..348S,2016ApJ...823..150T}
\begin{equation}
M_f\approx \frac{V_{\rm out}}{C_s}\simeq 5.8 l_{\rm loop,9.5}^{-1/7}B_{\rm rec,2}^{4/7}n_{\rm rec,9}^{-3/7}\propto l_{\rm loop}^{-2/7}\beta_{\rm rec}^{-2/7}n_{\rm rec}^{-1/7},\label{eq:Mach}
\end{equation}
where $C_s\approx \sqrt{\gamma k_BT_{\rm out}/m_p}$ is the sound speed, $\gamma=5/3$ is the specific heat ratio, $k_B$ is the Boltzmann constant, and $\beta_{\rm rec}$ is the plasma beta for the pre-reconnection plasma.
The values of $M_f$ for typical parameters in solar flares are high enough to form a collisionless shock \citep{2009A&ARv..17..409T} and accelerate particles  \citep{2014ApJ...780..125V}.

In solar flares, the energy released around the shock is transported to the chromosphere by non-thermal electron beams and thermal conduction, which evaporates the chromospheric plasma. The evaporation plasma fills the flare loop, whose temperature is estimated based on MHD simulations with thermal conduction \citep{2001ApJ...549.1160Y}:
\begin{equation}
T_{\rm evap}\approx \frac{T_{\rm out}}{3}\simeq3.4\times10^7 l_{\rm loop,9.5}^{2/7}B_{\rm rec,2}^{6/7}n_{\rm rec,9}^{-1/7}\rm~K,  
\end{equation}
where the factor of 1/3 comes from the numerical result.
The density of the evaporation plasma can be estimated by the balance between the magnetic pressure and the gas pressure \citep{2002ApJ...577..422S}:
\begin{equation}
 n_{\rm evap}\approx\frac{B_{\rm rec}^2}{16\pi k_BT_{\rm evap}}\simeq4.2\times10^{10} B_{\rm rec,2}^{8/7}l_{\rm loop,9.5}^{-2/7}n_{\rm rec,9}^{1/7}\rm~cm^{-3}.
\end{equation}

The evaporation plasma emits soft X-rays by thermal bremsstrahlung as observed in the soft X-ray band. The thermal free-free luminosity from the evaporation plasma is estimated to be \citep{rl79} 
\begin{eqnarray}
L_{\rm ff}&\simeq& 1.7\times10^{-27}\times f_{\rm vol}l_{\rm nl}l_{\rm loop}^2n_{\rm evap}^2T_{\rm evap}^{1/2}\rm~erg~s^{-1}\\
&\simeq&5.6\times10^{26}l_{\rm loop,9.5}^{18/7}B_{\rm rec,2}^{19/7}n_{\rm rec,9}^{3/14}f_{\rm vol,-0.5}f_{\rm nl,0.5} \rm~erg~s^{-1},\nonumber
\end{eqnarray}
where $f_{\rm vol}$ is the volume filling factor of the evaporation plasma, $l_{\rm nl}$ is the length of the magnetic neutral line, and $f_{\rm nl}=l_{\rm nl}/l_{\rm loop}$ (see Figure \ref{fig:schematic_sun}).
This value of $L_{\rm ff}$ corresponds to an X-class flare, with which GeV gamma rays are most frequently accompanied.
The cooling timescale of the evaporation plasma is estimated to be 
\begin{eqnarray}
t_{\rm ff}&=&\frac32\frac{n_{\rm evap}k_BT_{\rm evap}f_{\rm vol}l_{\rm loop}^2l_{\rm nl}}{L_{\rm ff}}\\
&\simeq&1.7\times10^4l_{\rm loop,9.5}^{1/7}B_{\rm rec,2}^{-5/7}n_{\rm rec,9}^{-3/14} \rm~s.\nonumber
\end{eqnarray}
The evaporation plasma falls back to the stellar surface in $t_{\rm ff}$, and this timescale corresponds to the duration of the gradual phase, i.e., $t_{\rm grad}\approx t_{\rm ff}$.

The evaporation plasma will fill a large portion of the reconnected magnetic loop within the travel timescale of the reconnection outflow. The chromospheric evaporation starts to occur in a timescale of non-thermal electron traveling across the loop, $t_{\rm nte}\approx l_{\rm loop}/c\simeq 0.1l_{\rm loop,9.5}\rm~s$, or the conduction cooling timescale, $t_{\rm cond}\approx {\rm max}(t_{\rm nte},~n_{\rm rec}k_Bl_{\rm loop}^2/(\kappa_0 T_{\rm out}^{5/2}))$. With our reference parameters, $t_{\rm cond}\simeq t_{\rm nte}$ is satisfied. These timescales are much shorter than the outflow travel timescale, 
\begin{equation}
    t_{\rm trav}=\frac{l_{\rm rec}}{V_{\rm out}}=9.2 l_{\rm loop,9.5}B_{\rm rec,2}^{-1}n_{\rm rec,9}^{1/2}f_{h,0.3}\rm~s,
\end{equation}
where $l_{\rm rec}$ is the position of the reconnection point above the loop top and $f_h=l_{\rm rec}/l_{\rm loop}$. Thus, we can regard that the chromospheric evaporation occurs immediately after the magnetic reconnection. 
The timescale for the evaporation plasma to fill the loop is estimated as 
\begin{equation}
    t_{\rm evap} = \frac{l_{\rm loop}}{C_s}=46 l_{\rm loop,9.5}^{6/7}B_{\rm rec,2}^{-3/7}n_{\rm rec,9}^{1/14}~{\rm s},
\end{equation}
where we approximate the upflow speed of the evaporation plasma to be the sound speed corresponding to the temperature of $T_{\rm evap}$. 
The ratio of $t_{\rm evap}$ to $t_{\rm trav}$ can be written as
\begin{equation}
    \frac{t_{\rm trav}}{t_{\rm evap}}\simeq 0.20 l_{\rm loop,9.5}^{-1/7}B_{\rm rec,2}^{4/7}n_{\rm rec,9}^{-3/7}f_{h,0.3}.
\end{equation}
Therefore, we expect that a non-negligible portion of the reconnected field line is filled with the evaporation plasma when this reconnected field passes through the termination shock. This picture is consistent with 2-D MHD simulations \citep{2015ApJ...805..135T}. Protons accelerated at the termination shock will interact with the evaporation plasma as discussed in Section \ref{sec:sun_had}.

Based on this scenario, the total released energy by reconnection is 
\begin{equation}
E_{\rm tot}\approx\frac{ l_{\rm loop}^2l_{\rm nl}B_{\rm rec}^2}{8\pi}\simeq4.0\times10^{31}l_{\rm loop,9.5}^3B_{\rm rec,2}^2f_{\rm nl,0.5}\rm~erg.
\end{equation}
Changing $B_{\rm rec}$, $l_{\rm loop}$, and $n_{\rm rec}$, we can reproduce the relation between the released energy and M- and X-class flares given in \citet{2015EP&S...67...59M}.
The total thermal energy of the evaporation plasma is 
\begin{eqnarray}
 E_{\rm evap}&\approx& l_{\rm loop}^2l_{\rm nl}\frac32n_{\rm evap}k_BT_{\rm evap}\\
&\simeq&9.4\times10^{30} f_{\rm vol,-0.5}l_{\rm loop,9.5}^3B_{\rm rec,2}^2f_{\rm nl,0.5}\rm~erg. \nonumber
\end{eqnarray}
Since the parameter dependence of these two energies are the same except for $f_{\rm vol}$, $E_{\rm tot}>E_{\rm evap}$ is satisfied as long as we choose $f_{\rm vol}<1$.

\subsection{Hadronic emission}\label{sec:sun_had}

Here, we consider two processes in two different zones: particle acceleration at the termination shock and gamma-ray emission in the flare loops. 
To obtain the gamma-ray spectrum of a solar flare, we consider the transport equation for CRs in the flare loops with the one-zone approximation:
\begin{equation}
 \frac{\partial N_{E_p}}{\partial t}-\frac{\partial}{\partial E_p}\left(\frac{E_p N_{E_p}}{t_{\rm cool}}\right)=-\frac{N_{E_p}}{t_{\rm esc}}+\dot{N}_{E_p,\rm inj},\label{eq:transport}
\end{equation}
where $E_p$ is the proton energy, $N_{E_p}=dN/dE_p$ is the number spectrum for CR protons, $t_{\rm cool}$ is the cooling timescale, $t_{\rm esc}$ is the escape time from the radiation region (i.e., the escape timescale from the flare loop), and $\dot{N}_{E_p,\rm inj}$ is the injection term. We explain the individual term in the following paragraphs. 

First, we describe the injection term. We consider that CR particles are accelerated at the termination shock and advected to the flare loop by turbulent motion. 
The acceleration time is given by \citep[e.g.,][]{Dru83a}, 
\begin{equation}
t_{\rm acc}\approx \frac{20\xi E_p}{3eB_{\rm rec}c}\left(\frac{c}{V_A}\right)^2,
\end{equation}
where $\xi$ is the Bohm factor (physically, the mean free path of protons divided by the gyro radius).
The advection timescale can be evaluated using the shock width and the outflow velocity to be
\begin{equation}
t_{\rm sh}\approx \frac{l_{\rm out}}{V_{\rm adv}}\simeq0.14l_{\rm loop,9.5}B_{\rm rec,2}^{-1}n_{\rm rec,9}^{1/2}\eta_{\rm rec,-1.5}f_{\rm adv}^{-1} \rm~sec,
\end{equation}
where $V_{\rm adv}=f_{\rm adv}V_A$ is the advection velocity at the shock downstream and $f_{\rm adv}$ is a parameter.
The diffusive shock acceleration process accelerates CR protons with a power-law energy distribution below the maximum energy given by balancing $t_{\rm sh}$ and $t_{\rm acc}$. The maximum energy is estimated to be
\begin{eqnarray}
 E_{p,\rm max}&\approx& \frac{3eB_{\rm rec}V_Al_{\rm out}}{20\xi' c}\label{eq:Epmax}\\
&\simeq& 5.1 B_{\rm rec,2}^2n_{\rm rec,9}^{-1/2}l_{\rm loop,9.5}\eta_{\rm rec,-1.5}{\xi'}^{-1}_{0.3}\rm~GeV,\nonumber
\end{eqnarray}
where $\xi'=\xi f_{\rm adv}$. Here, we give $\xi'$ as a parameter.
This energy is higher than the pion production threshold, $E_{p,\rm thr}\simeq1.22$ GeV, and thus, these CRs can emit high-energy gamma rays via hadronuclear interactions.
In reality, multiple termination shocks may exist above the loop top region \citep{2016ApJ...823..150T,2022arXiv221205802S}, but we consider the CR particle acceleration at a single shock for simplicity.
In addition, stochastic acceleration by turbulence can be effective at the shock downstream, which may affect the gamma-ray emission during the flare \citep[e.g.,][]{2012SSRv..173..535P,2015ApJ...806...80K}. However, current particle-in-cell (PIC) simulations on proton acceleration at non-relativistic shocks do not show any evidence of stochastic acceleration, even though the wave-particle scattering is very efficient \citep[e.g.,][]{CS14a,CS14b,CS14c}. Here, we do not model the stochastic acceleration process in detail.

Since $t_{\rm sh}\ll t_{\rm rec}$ and $t_{\rm sh}\ll t_{\rm ff}$, we can regard that the CR particles are instantaneously injected to the flare loop with a power-law form of $\dot{N}_{E_p,\rm inj}\propto E_p^{-s}$. The spectral index is given by $s=(r+2)/(r-1)$, where $r=(\gamma+1)/(\gamma-1+2/M_f^2)$ is the compression ratio \citep[e.g.,][]{BE87a}. We obtain $r\simeq3.67$ and $s\simeq2.12$ with our reference parameters. We consider the sheet-like geometry of the reconnection outflows and give normalization of the injection term as
\begin{eqnarray}
&\int& \dot{N}_{E_p,\rm inj}dE_p=L_p= \epsilon_p n_{\rm rec}m_pV_A^3l_{\rm nl}l_{\rm out}\label{eq:Lp}\\
&\simeq& 5.5\times10^{26}\epsilon_{p,-3}l_{\rm loop,9.5}^2B_{\rm rec,2}^3n_{\rm rec,9}^{-1/2}\eta_{\rm rec,-1.5}f_{\rm nl,0.5}\rm~erg~s^{-1} \nonumber,
\end{eqnarray}
where $\epsilon_p$ is the production efficiency of relativistic CRs and $L_p$ is the CR proton luminosity.
The termination shock should be an oblique shock in which the magnetic field is almost perpendicular to the shock normal, as shown in MHD simulations  \citep{2015ApJ...805..135T,2016ApJ...823..150T,2018ApJ...869..116S}, but the values of $\epsilon_p$ in oblique shocks are uncertain.
Particle-in-cell simulations suggest that oblique shocks are likely inefficient to produce CRs \citep{CS14a,2015ApJ...798L..28C}. On the other hand, some theoretical models argue that oblique shocks can accelerate CR particles more efficiently than parallel shocks \citep{2022ApJ...925...48X}, and observations of supernova remnants support the efficient particle acceleration at perpendicular shocks \citep{2017A&A...597A.121W}.
In this paper, we tune $\epsilon_p$ so that the resulting gamma-ray spectrum matches the Fermi-LAT data.

Next, we explain the escape term. In the flare loops, the magnetic field strength at the footpoints is stronger than that in the loop-top region (see the inset of Figure \ref{fig:schematic_sun}). Thus, the flare loops work as a magnetic bottle that confines CRs efficiently \citep{1997ApJ...485..859S,1998ApJ...502..468A}. The CRs with a small pitch angle can escape from the flare loops, and thus, the timescale of the CR escape should be comparable to that of the pitch-angle scattering \citep[e.g.,][]{sp08,kmt15}: 
\begin{equation}
 t_{\rm esc}\approx \eta_{\rm turb}\frac{l_{\rm loop}}{c}\simeq 1.1\eta_{\rm turb,1}l_{\rm loop,9.5} \rm~s,\label{eq:tesc}
\end{equation} 
where $\eta_{\rm turb}=B^2/(8\pi\int \mathcal{P}_k dk)$ is the turbulence strength parameter, $\mathcal{P}_k$ is the turbulence power spectrum, and we assume Goldreich-Sridhar turbulence of $\mathcal{P}_k\propto k^{-2}$ \citep{gs95}. This turbulence power spectrum leads to the hard-sphere type scattering in which the scattering timescale is independent of the particle energy.
The value of $t_{\rm esc}$ might also depend on the wave-particle interaction rate \citep[e.g.,][]{1999ApJ...519..422K}. We should note that the assumed turbulence properties are different in the particle acceleration region (the termination shocks: Bohm-like turbulence) and gamma-ray emission region (the flare loops: Goldreich-Sridhar turbulence), and we introduce two different parameters to describe the turbulent strength, $\xi$ and $\eta_{\rm turb}$. 

Finally, we explain the cooling term. We consider only $pp$ inelastic collisions ($p+p\rightarrow p+p+\pi$), because other processes, such as photomeson production ($p+\gamma\rightarrow p+\pi$), Bethe-Heitler ($p+\gamma\rightarrow p+e^++e^-$), and proton synchrotron processes are negligible. Then, the cooling timescale is given  by $t_{\rm cool}\approx t_{pp}\approx1/(n_{\rm evap}\sigma_{pp}\kappa_{pp}c)$, where  $\sigma_{pp}\sim30$ mb and $\kappa_{pp}\simeq0.5$ are the crosssection and inelasticity for $pp$ inelastic collisions, respectively.
We consider that the flare loops do not expand with time, and thus, we can neglect the adiabatic expansion in the cooling term. Indeed, \cite{2015ApJ...805..135T} analyzed the compression and expansion of plasma around the flare loop top (see their Figures 4 and 6. See also Figure 4 of \citealt{2016ApJ...823..150T}). These simulations show that the adiabatic expansion is insignificant in most of the regions around the termination shocks and in the flare loop. The adiabatic expansion may be significant only in localized small regions around the termination shocks.


There are three relevant timescales for gamma-ray emission from solar flares. One is the reconnection timescale, $t_{\rm rec}$, which indicates the duration of CR production. Another is the CR escape timescale, $t_{\rm esc}$, after which CRs escape from the emission region. The other is the cooling timescale of the evaporation plasma, $t_{\rm ff}$, after which the target of hadronuclear interactions disappears. For luminous solar flares, $t_{\rm esc}<t_{\rm rec}<t_{\rm ff}$ is always satisfied.
Then, the CRs escape from the loop while they are continuously accelerated at the shock and provided into the loop. Thus,
the CR spectrum in the loop should be determined by the balance between the injection and escape, $N_{E_p}\approx \dot{N}_{E_p,\rm inj}t_{\rm esc}$, and the steady state is achieved.
The CR protons interact with the evaporation plasma and produce charged and neutral pions with the ratio of $\pi^\pm:\pi^0\sim2:1$. Charged pions decay to neutrinos and electrons/positrons, while neutral pions decay to gamma rays. The pion-decay gamma-ray spectra are roughly estimated to be \citep[e.g.,][]{mal13,2017PTEP.2017lA105A}
\begin{equation}
 E_\gamma L_{E_\gamma}\approx \frac{1}{3}f_{pp} E_p^2 \dot{N}_{E_p,\rm inj},
\end{equation}
where $E_\gamma\approx0.1E_p$ is the typical gamma-ray energy produced by pion decay and $f_{pp}={\rm min}(t_{\rm esc}/t_{pp},~1)\simeq2.4\times10^{-5}l_{\rm loop,9.5}^{5/7}B_{\rm rec,2}^{8/7}n_{\rm rec,9.5}^{1/7}\eta_{\rm turb,1}$ is the pion production efficiency. Since the CR spectrum is soft ($s>2$) for solar flares, the gamma-ray spectrum has a peak around the pion production threshold, $E_{\gamma,\rm pk}\sim100$ MeV. The peak luminosity of the pion-decay gamma rays is estimated to be
\begin{eqnarray}
& & L_{\gamma,\rm pk}\approx \frac13f_{pp}L_p\simeq3.6\times10^{21} \label{eq:Lgam}\\
&\times& \epsilon_{p,-3}l_{\rm loop,9.5}^{19/7}B_{\rm rec,2}^{29/7}n_{\rm rec,9}^{-5/14}\eta_{\rm rec,-1.5}\eta_{\rm turb,1}f_{\rm nl,0.5}\rm~erg~s^{-1}.\nonumber
\end{eqnarray}
The estimated gamma-ray luminosity with $\epsilon_p=10^{-5}-10^{-2}$ is consistent with the Fermi-LAT observations of the impulsive phase \citep{2021ApJS..252...13A}.

The pion decay process also produces secondary neutrinos and electron-positron pairs. We cannot expect detection of neutrinos in GeV energies because of the small neutrino-nucleon interaction crosssection and strong atmospheric background. The luminosity of the secondary electron-positron pairs is comparable to the luminosity of gamma rays, which is likely much lower than the primary non-thermal electrons accelerated together with protons (see Section \ref{sec:leptonic}). Therefore, we do not discuss the emission by the secondary electron-positron pairs.

To quantitatively compare our model to the observations, we numerically calculate the hadronic gamma-ray spectrum from solar flares. We use the delta-function approximation \citep{2000A&A...362..937A} with $\sigma_{pp}$ calibrated by the LHC data \citep{2014PhRvD..90l3014K}. The proton spectrum is obtained by solving Equation (\ref{eq:transport}) with a steady state approximation \cite[see, e.g.,][]{2019PhRvD.100h3014K,2020ApJ...904..188K,2020ApJ...905..178K}.

\begin{figure}
   \begin{center}
    \includegraphics[width=\linewidth]{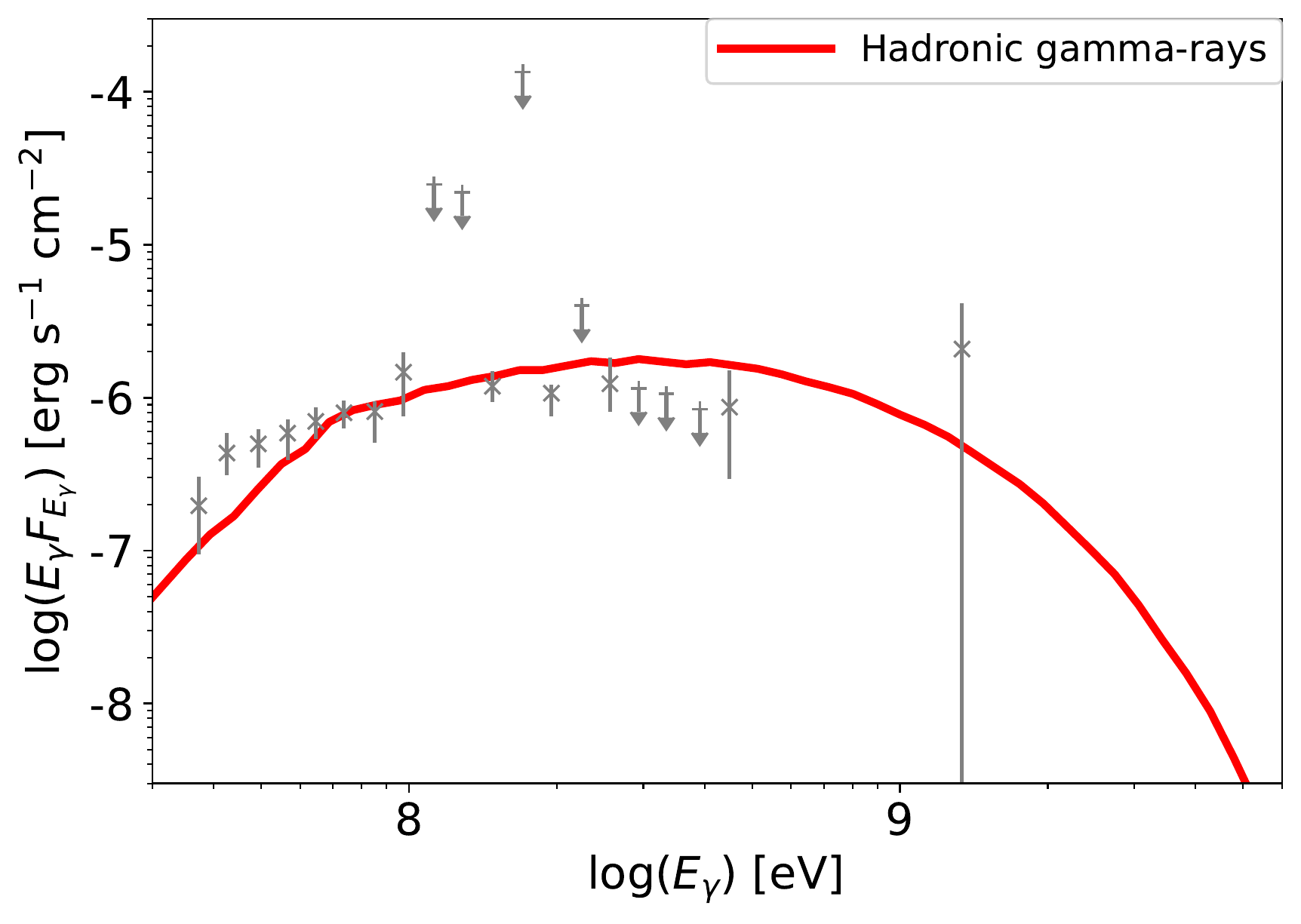}
    \caption{High-energy gamma-ray spectra in the impulsive phase of the solar flare on 2014 February 25. The red line is our model prediction, and the grey simbols with error bars are the Fermi-LAT data \citep{2021ApJS..252...13A}. }
    \label{fig:solar}
   \end{center}
  \end{figure}

Figure \ref{fig:solar} exhibits the gamma-ray spectra for our model and for the impulsive phase of a Fermi-LAT solar flare on 2014 February 25 \citep{2021ApJS..252...13A}. We tabulate our model parameters and resulting quantities in Table \ref{tab:results}.  Our parameter choice leads to an X-class flare (X2.7), in rough agreement with the observed class (X4.9). Our model can roughly reproduce the observed gamma-ray spectrum with a reasonable parameter set. \citet{2014ApJ...797L..15C} estimated the loop size and the density of the X-ray emission region, and our parameter choice is in rough agreement with theirs. Also, \citet{2017ApJ...835..139S} estimated the reconnection rate in the gradual phase to be $4\times10^{-3}-7\times10^{-3}$, which is an order of magnitude lower than that used in our calculations. However, the reconnection rate in the impulsive phase can be higher than that in the gradual phase \citep[e.g.,][]{2012ApJ...745L...6T,2022SoPh..297...80Q}.
Therefore, we consider that our parameter choice does not contradict with that by \cite{2017ApJ...835..139S}.

\begin{table*}
\begin{center}
\caption{List of parameters and resulting physical quantities for solar and protostellar flares. Parameters for the solar flare are calibrated using the data for the event on 2014 February 25. The values for Model C  are identical to those for Model B, except for $\epsilon_p$ and $L_p$.} 
\label{tab:results} 
 
\begin{tabular}{|c|cccccccccc|}
\hline
 Given parameters  & & & & & & & & & & \\
\hline
Model & $l_{\rm loop}$ & $B_{\rm rec}$ & $n_{\rm rec}$ & $f_{\rm nl}$ & $\epsilon_p$ & $\eta_{\rm rec}$ & $f_{\rm vol}$ & $\eta_{\rm turb}$ & $\xi$ & $\epsilon_e/\epsilon_p$ \\ 
 & [$10^{10}$ cm] & [kG] & [$10^9$ cm$^{-3}$] & & & & & & & \\
\hline
Sun & 0.32 & 0.10 & 1.0 & 6.0 & 0.004 & 0.032 & 0.32 & 10 & 2 & 0.032 \\
Model A & 10 & 1.0 & 10 & 1.0 & 0.004 & 0.032 & 0.32 & 10 & 2 & 0.032 \\
Model B & 13 & 2.0 & 10 & 1.0 & 0.004 & 0.032 & 0.32 & 10 & 2 & 0.032 \\
Model C & 13 & 2.0 & 10 & 1.0  & 0.1 & 0.032 & 0.32 & 10 & 2 & 0.032 \\
\hline
\end{tabular}
 
\begin{tabular}{|c|cccccccccc|}
\hline
 Resulting quantities & & & & & & & & & & \\
\hline
Model & $V_A$ & $M_f$ & $s$ & $\log_{10}(L_{\rm ff})$ & $T_{\rm evap}$ & $t_X$ & $t_\gamma$ & $\log_{10}(E_{p,\rm max})$ & $\log_{10}(L_p)$ & $\log_{10}(\nu_{\rm ssa})$  \\
      & [$10^8\rm~cm~s^{-1}$]  &  &   & [erg s$^{-1}$]    &  [$10^7$K]          & [hr]   &   [min] & [GeV] & [erg s$^{-1}$] & [Hz]\\
\hline
Solar Flare & 6.9 & 5.8 & 2.12 & 27.02 & 3.4 & 4.7 & 2.4 & 0.71 & 27.62 & 8.93 \\
Model A & 21.8 & 4.9 & 2.17 & 33.03 & 47.5 & 2.4 & 24.2 & 3.71 & 32.34 & 10.76 \\
Model B & 43.7 & 7.1 & 2.08 & 34.14 & 92.7 & 1.7 & 15.7 & 4.43 & 33.47 & 11.28 \\
Model C & 43.7 & 7.1 & 2.08 & 34.14 & 92.7 & 1.7 & 15.7 & 4.43 & 34.87 & 11.67 \\
\hline
\end{tabular} 
\end{center}
\end{table*}

\section{Application to protostellar flares}\label{sec:protostar}

  \begin{figure*}
   \begin{center}
    \includegraphics[width=\linewidth]{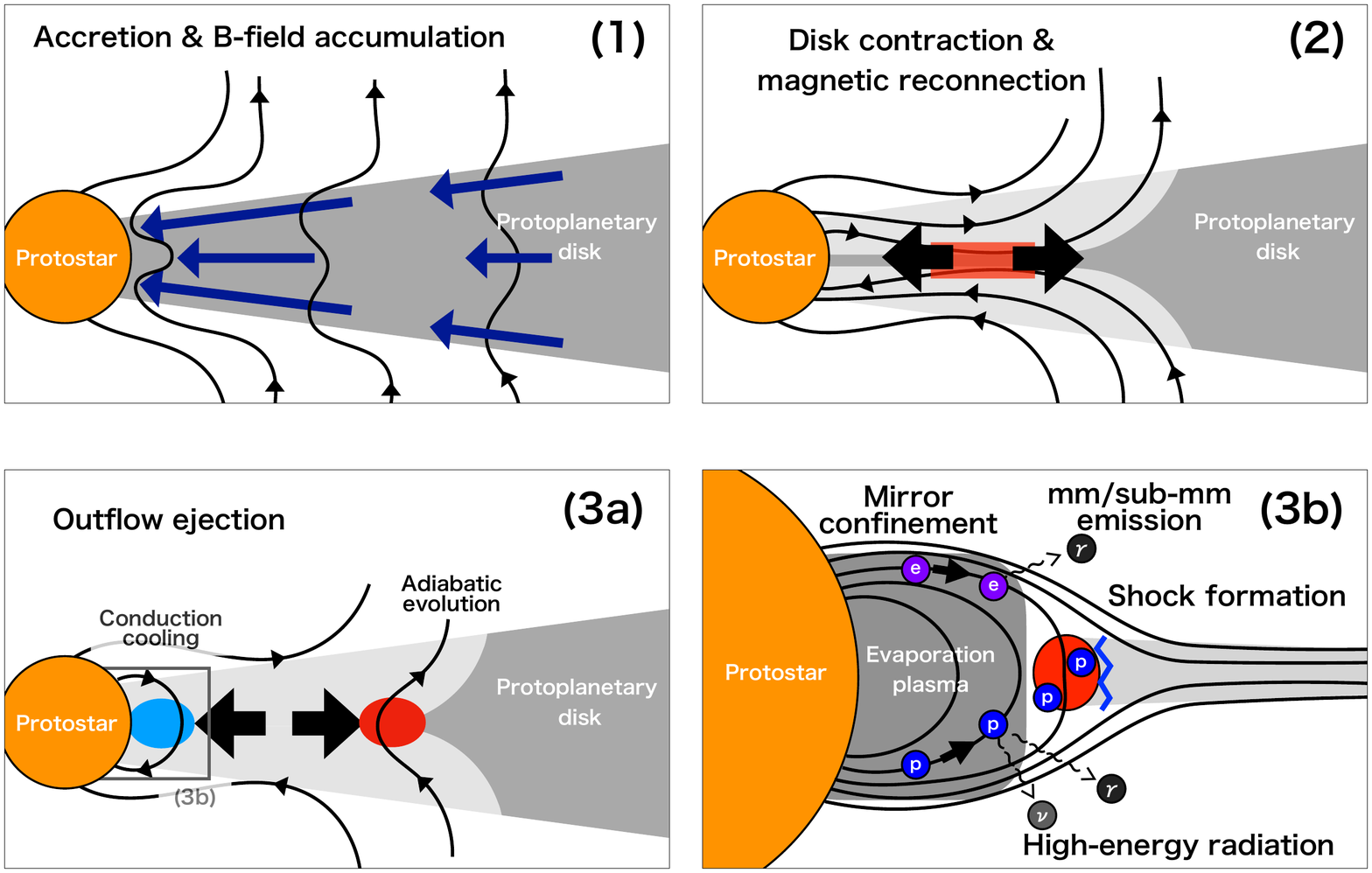}
    \caption{Schematic picture of our scenario of gamma-ray flares from protostars. Panel (1): Mass accretion from a protoplanetary disk accumulates magnetic fields around a protostar. The surface layers of the disk accrete faster than the midplane. This flow structure stretches the magnetic fields. Panel (2): When magnetic field strength becomes comparable to the thermal energy of the disk gas, magnetic fields and the coronal plasma starts to expand toward the disk midplane, which clears out the disk. This  eventually triggers magnetic reconnection. Panel (3a): The magnetic reconnection produces bipolar outflows. One moves to the protostar, and it cools down by thermal conduction before colliding with the flare loop.  The other outflow moves to the protoplanetary disk, and it evolves adiabatically. Panel (3b): Thermal conduction to the protostellar surface leads to evaporation of the protostellar atmosphere, and the evaporation plasma emits X-rays via thermal bremsstrahlung. Also, the collision between the outflow and the flare loop forms a shock in the outflow, where non-thermal particles are accelerated via the diffusive shock acceleration process. The accelerated protons emit gamma rays and neutrinos via hadronuclear interactions with evaporation plasma, and non-thermal electrons emit mm/sub-mm photons by synchrotron radiation.  }
    \label{fig:schematic}
   \end{center}
  \end{figure*}

\subsection{The accretion-driven flare scenario}

We apply our model to protostellar flares. Our discussion on protostellar flares is based on the scenario proposed by \cite{2019ApJ...878L..10T}, which is summarized in Figure \ref{fig:schematic}. They focus on the efficient magnetic transport driven by the disk surface accretion \cite[coronal accretion:][]{1996ApJ...461..115M,2009ApJ...707..428B,2018ApJ...857....4T}. The accretion flows carry disk poloidal fields to the protostar, and the large-scale fields are accumulated (panel 1). 
When the magnetic pressure around the protostar becomes comparable to the disk gas pressure, the protostellar field starts to expand toward the equator. The expanding field interacts with the disk surfaces and exchanges the angular momentum. This reduces the disk surface density around the interacting region, and the protostellar fields expand further from the northern and southern hemispheres. Finally, the protostellar fields reconnect each other around the equatorial plane, which drives a flare (panel 2). 
The magnetic reconnection releases sufficient magnetic energy to account for powerful X-ray flares with $E_X\sim10^{37}$ erg.

A magnetic reconnection drives bipolar outflows; one moving toward the protostar and the other heading to the protoplanetary disk (panel 3a). 
The outflow heading toward the protostar forms the closed magnetic arcade or flare loop. We can regard this closed flare loop as the scale-up version of a solar flare loop.  
Thus, we can apply our solar flare model to this system (panel 3b), although the triggering mechanism is different from stellar flares. 
We note that the triggering mechanism of X-ray flares in the more evolved YSOs could be similar to that in the main sequence stars including the Sun \citep[e.g.][]{2021ApJ...916...32G,2021ApJ...920..154G}. 
The other outflow colliding with the disk evolves nearly adiabatically because the magnetic field accompanied by the hot plasma is connected to neither the protostellar surface nor the disk gas.
In this case, the acoustic Mach number is close to 1 (see Equations (\ref{eq:Tout}) and (\ref{eq:Mach})), and thus, we cannot expect particle acceleration at the termination shock.

\subsection{Physical quantities around the reconnection region}

First, we estimate the physical quantities of the magnetic reconnection region before the reconnection. As a reference parameter set, we take the protostellar mass $M_*=0.5M_\odot$ and radius $R_*=2R_\odot$. The magnetic reconnection occurs at a radius of $R_{\rm dis}\sim 2R_*$. 
The resulting flare loop will have a size similar to the protostar, i.e., $l_{\rm loop}\sim10^{11}$ cm.

As pointed out by \cite{2019ApJ...878L..10T}, the stellar field strength before a flare is determined by the balance between the magnetic pressure around the poles and the disk gas pressure. This condition is similar to the magnetically arrested disk (MAD) state in black hole accretion \citep{br74,NIA03a,MTB12a}. 
We first estimate the disk gas pressure around $R_{\rm dis}$, and then, calculate the reconnection field strength, $B_{\rm rec}$. 
The disk density at $R\sim R_{\rm dis}$ can be estimated using the alpha-viscosity prescription:
\begin{eqnarray}
& &n_d\approx \frac{\dot{M}}{4\pi R_{\rm dis}^2 \alpha\mathcal{H}^3V_K}\label{eq:nd}\\
& &\simeq2.1\times10^{16}\dot{M}_{-6.5}M_{-0.3}^{-1/2}R_{\rm dis,0.3}^{-3/2}\mathcal{H}_{-1}^{-3}\alpha_{-1}\rm~g~cm^{-3},\nonumber
\end{eqnarray}
where $V_K=\sqrt{GM/R_{\rm dis}}\simeq2.2\times10^7 M_0 R_{\rm dis,0.3}\rm~cm~s^{-1}$ is the Keplerian velocity, $\mathcal{H}=H/R_{\rm dis}\approx C_s/V_K\sim0.1$ is the disk aspect ratio, $M$ is the mass of the protostar, $H$ is the pressure scale height, $C_s$ is the sound speed in the protoplanetary disk, $R_{\rm dis,0.3}=R/(2R_\odot)$, $\dot{M}_{-6.5}=\dot{M}/(10^{-6.5}~M_\odot\rm~yr^{-1})$, and $M_{-0.3}=M/(0.5~M_\odot)$. 
When the magnetic reconnection occurs, the magnetic pressure around the protostar is comparable to the disk gas pressure\footnote{In black hole accretion flows, we use the critical magnetic flux, $\Phi_{\rm MAD}=2\pi R_G B_{\rm rec}/\sqrt{\dot{M}c}\sim50$, to estimate the magnetic fields around black holes \citep[e.g.,][]{YN14a,2022ApJ...937L..34K}. The order of $\Phi_{\rm MAD}$ is obtained by balancing the magnetic energy and gravitational energy, which is different from Equation (\ref{eq:Brec}). }.
The magnetic field strength at the magnetic reconnection point is estimated to be
\begin{equation}
 B_{\rm rec}\approx\sqrt{8\pi n_d m_pC_s^2}\simeq 1.4\times10^3n_{d,16}^{1/2} M_{-0.3}^{1/2} R_{\rm dis,0.3}^{-1/2}\mathcal{H}_{-1}\rm~G. \label{eq:Brec}
\end{equation}

The magnetic reconnection occurs when the disk matter is cleared (see Figure~\ref{fig:schematic}).
At this stage, the stellar coronal gas expands toward the reconnection point near the equatorial plane. Therefore, the density just around the reconnection region is the coronal value and should be much smaller than the disk density.
We assume $n_{\rm rec}=10^{10}\rm~cm^{-3}$ as a fiducial value, which is higher than the solar coronal value ($\sim10^8-10^9\rm~cm^{-3}$), because of the following reasons. First, protostars typically exhibit stronger magnetic activities than the Sun. Second, more X-ray luminous stars have larger magnetic fluxes \citep{2003ApJ...598.1387P,2020ApJ...901...70T}. Third, the coronal density in protostars could be enhanced by accretion heating \citep{2008ApJ...689..316C}.
In the rest of this paper, we provide $l_{\rm loop}$, $n_{\rm rec}$, and $B_{\rm rec}$ as primary parameters for the magnetic reconnection events so that we can apply the scenario in the previous section.

\subsection{Thermal emissions from evaporation plasma}

X-ray observations revealed that many protostars
exhibit luminous X-ray flares \citep{2001ApJ...557..747I,2008ApJ...677..401P}. 
We search for suitable parameter sets within the range of $B_{\rm rec}=3\times10^2-3\times10^3$ G, $l_{\rm loop}=3\times10^{10}-3\times10^{11}$ cm, and $n_{\rm rec}=10^{10}-10^{11}\rm~cm^{-3}$ so that our model results are consistent with X-ray observations summarized below.
The X-ray luminosity, duration, and temperature of protostellar flares lie in the following ranges:  $3\times10^{30}{\rm~erg~s^{-1}}\lesssim L_X\lesssim10^{34}{\rm~erg~s^{-1}}$, $3\times10^4{\rm~s}\lesssim t_X\lesssim10^5$ s, $10^7{\rm~K}\lesssim T_{\rm evap}\lesssim10^9$ K \citep{2003PASJ...55..653I,2008ApJ...688..418G,2021ApJ...916...32G,2021ApJ...920..154G}, where $L_X$ is evaluated in the Chandra band (0.5 keV -- 7 keV).
MAXI provides an additional constraint on X-ray luminosity.
MAXI does not detect any flares from protostars in Taurus Molecular Cloud and Rho Ophuchi Cloud Complex \citep{2016PASJ...68...90T}. Since the cadence of the MAXI observation (once in a 90 min) is shorter than the typical duration of protostellar flares, non-detection means that X-ray emissions from protostellar flares should be lower than the MAXI sensitivity ($\sim10^{-9}\rm~erg~s^{-1}~cm^{-2}$ for the $2-20$ keV band; \citealt{2009PASJ...61..999M}).

\begin{figure*}
   \begin{center}
    \includegraphics[width=\linewidth]{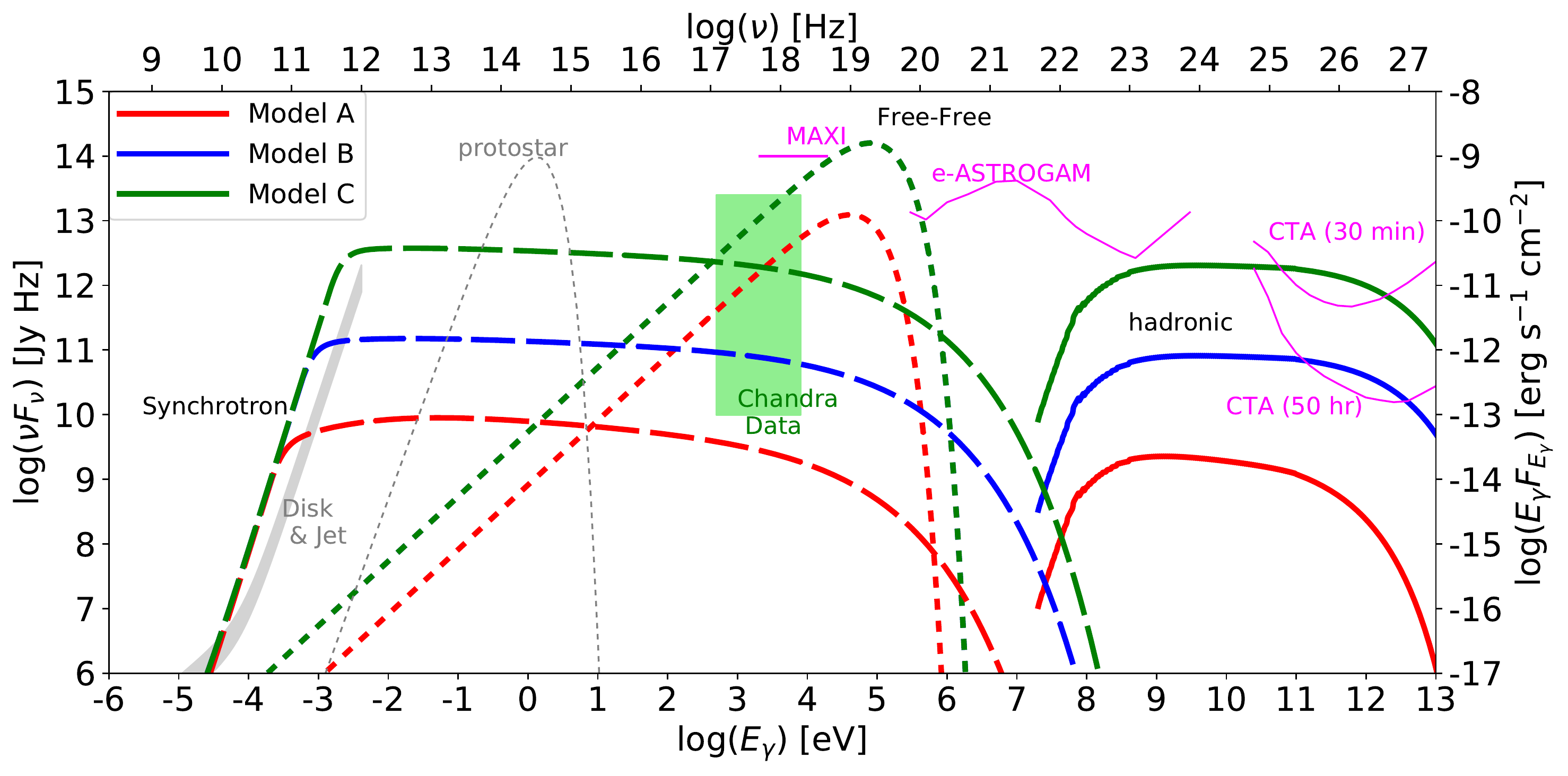}
    \caption{Broadband photon spectra from protostellar flares. The thick-solid, thick-dashed, and thick dotted lines show the hadronic pion-decay, leptonic synchrotron, and thermal free-free components, respectively. The red, blue, and green lines are for Model A, B, and C, respectively. The thermal free-free emission for Model B is identical to that for Model C.  The gray-shaded region and grey-thin-dotted line show the disk \& Jet components given in \citet{2012MNRAS.420.3334S} and the thermal spectrum from the protostar with $R_*=2R_\odot$ and $T=4000$ K, respectively. The lightgreen-shaded region shows the flux range of protostellar flare observations by Chandra \citep{2008ApJ...688..418G,2021ApJ...920..154G}. The magenta-thin-solid lines show the sensitivity curves for MAXI \citep{2009PASJ...61..999M}, e-ASTROGAM \citep{2017ExA....44...25D}, and CTA \citep{2019scta.book.....C}. The thermal free-free components are consistent with the Chandra observations. CTA will be able to detect the hadronic emission for model C. The leptonic synchrotron emission can be detected by mm/sub-mm surveys.}
    \label{fig:allband}
   \end{center}
  \end{figure*}

Table \ref{tab:results} lists the parameter sets whose resulting X-ray fluxes are consistent with the Chandra and MAXI data.  
Figure \ref{fig:allband} shows photon spectra by free-free emission from thermal electrons (see thick-dotted lines) for our protostellar flare models A (red), B (blue), and C (green). 
The thick dotted lines for models B and C are completely overlapped. 
The durations of X-ray flares are a few hours, and the temperature of evaporation plasma is $5\times10^8-9\times10^9$ K. The thermal free-free flux below the UV range is lower than those from other components, such as a protostar (thin-dotted line), a protoplanetary disk (gray-shaded region), and protostellar jets (gray-shaded region).

\subsection{Hadronic gamma-ray emissions}

We discuss non-thermal particle acceleration and emissions in protostellar flares. In order to accelerate particles at the shock, the optical depth for the shock upstream should be optically thin \citep[e.g.,][]{mi13,KMB18a}. Otherwise, the photons diffusing from the shock downstream decelerate the upstream fluid significantly. This causes a gradual velocity change at the shock \citep{BKS10a,2020PhR...866....1L}, and CR particles are no longer able to cross the shock. In the pre-shock outflow, the opacity is dominated by the electron scattering \citep{BL94a}. The optical depth for the reconnection outflow is estimated to be $\tau_T\approx n_{\rm rec}\sigma_Tl_{\rm loop}\simeq7\times10^{-4}n_{\rm rec,10}R_{\rm rec,11}$, and thus, the radiation mediated condition is easily avoided with our reference parameters\footnote{Accretion shocks at the surface of protostars are discussed as a CR production site \citep[e.g.,][]{pad16}, but the accretion shocks can be optically thick because the density at the accretion shock can be as high as $n_d$ given in Equation (\ref{eq:nd}). We need to check the radiation-mediated condition when discussing the non-thermal processes in dense environments, which is often ignored in the previous literature.}.

  \begin{figure}
   \begin{center}
    \includegraphics[width=\linewidth]{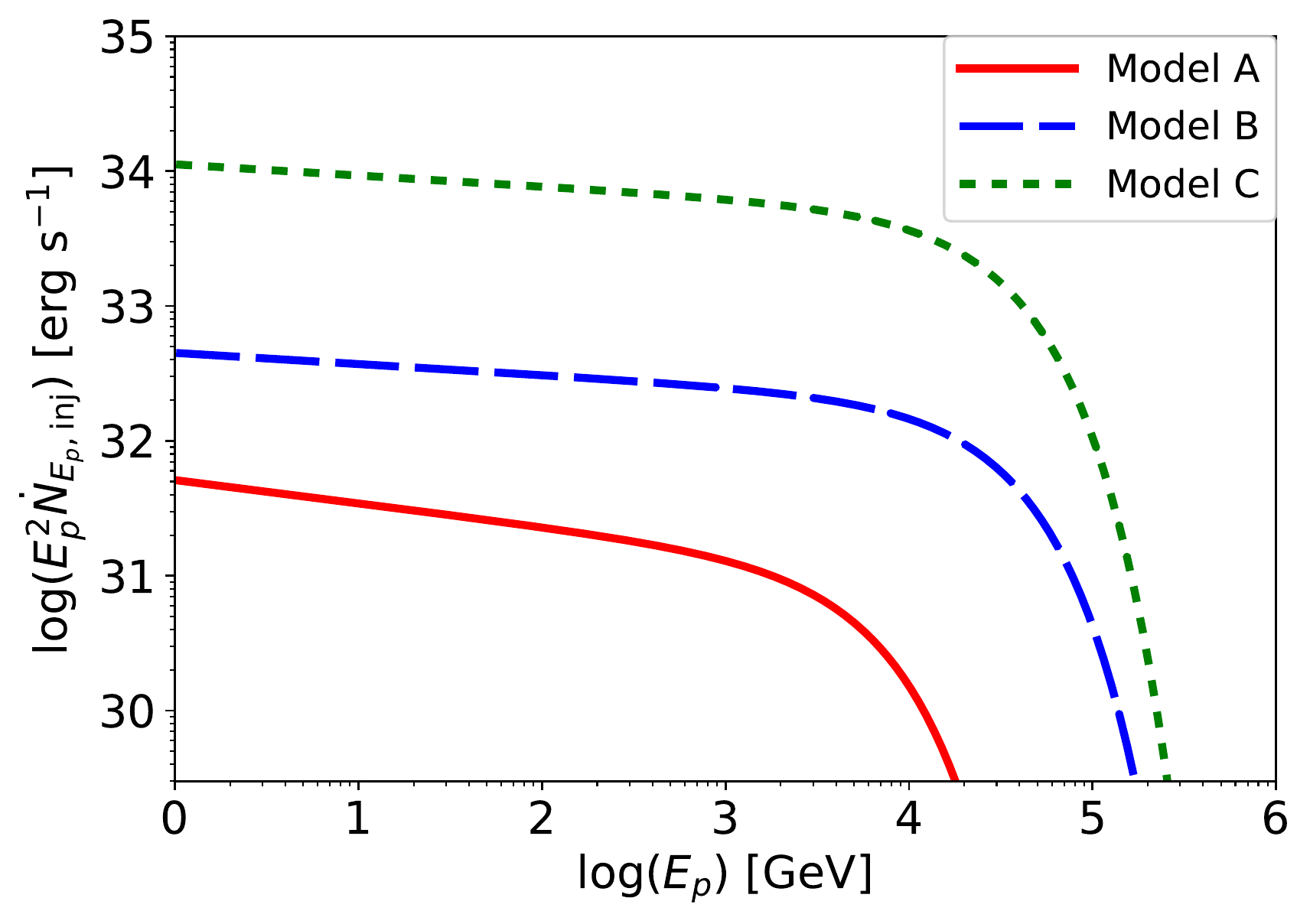}
    \caption{Spectra of injected non-thermal protons. The red-solid, blue-dashed, and green-dotted lines are for models A, B, and C, respectively.}
    \label{fig:crspectrum}
   \end{center}
  \end{figure}

  In the range of our investigation of protostellar flares, $t_{\rm sh}<t_{\rm esc}<t_{\rm rec}<t_{\rm ff}$ is satisfied. This situation is the same as that for solar flares, and thus, we can apply the model discussed in Section \ref{sec:sun_had}. We find that photomeson production, Bethe-Heitler, and proton synchrotron processes are negligible even with the parameters for protostellar flares. 
  
  According to Equations (\ref{eq:Epmax}) and (\ref{eq:Lp}), $L_p$ and $E_{p,\rm max}$ strongly depend on $l_{\rm loop}$ and $B_{\rm rec}$. Thus, the protostellar flares can produce cosmic rays and gamma rays much more efficiently, owing to their large loop size and strong magnetic field. We find that the protostellar flares can accelerate protons up to multi-TeV range, as shown in Figure \ref{fig:crspectrum} (see Table \ref{tab:results} for the values of $E_{p,\rm max}$). These protons can produce very-high-energy (VHE: $E_\gamma>100$ GeV) gamma rays, where the sensitivity to a transient source is better than the lower energy gamma-ray bands.

The gamma-ray luminosity at the VHE range is analytically estimated to be
\begin{eqnarray}
 L_{\gamma,\rm vhe}&\approx&\frac13f_{pp}f_{\rm vhe}L_p\simeq1.7\times10^{29} l_{\rm loop,11}^{19/7}B_{\rm rec,3}^{29/7}n_{\rm rec,10}^{-5/14}\nonumber\\
&\times& \epsilon_{p,-3}\eta_{\rm rec,-1.5}\eta_{\rm turb,1}f_{\rm nl}\left(\frac{f_{\rm vhe}}{0.067}\right)\rm~erg~s^{-1},
\end{eqnarray}
where $f_{\rm vhe}=({\rm TeV}/m_pc^2)^{2-s}(s-2)/(1-(E_{p,\rm max}/m_pc^2)^{2-s})$ is the correction factor from the bolometric luminosity to the VHE range \footnote{$f_{\rm vhe}$ is computed as follows. We use Eq. (\ref{eq:Lp}) to obtain the normalization of the CR proton spectrum. The CR proton spectrum has a peak at $E_p\approx m_pc^2$ for $s>2$. VHE gamma rays are produced by CR protons of $E_p\gtrsim1$ TeV, but the number density of protons at this energy is lower than that at $E_p\sim m_pc^2$. We take into account such a reduction of the number of CR protons when computing $f_{\rm vhe}$.}. The VHE flux is estimated to be
\begin{eqnarray}
 & &F_{\gamma,\rm vhe}=\frac{L_{\gamma,\rm vhe}}{4\pi d^2}\simeq 4.6\times10^{-15}  l_{\rm loop,11}^{19/7}B_{\rm rec,3}^{29/7}n_{\rm rec,10}^{-5/14}\epsilon_{p,-3}\nonumber\\
&\times& \eta_{\rm rec,-1.5}\eta_{\rm turb,1}f_{\rm nl}\left(\frac{f_{\rm vhe}}{0.065}\right)\left(\frac{d}{140\rm~pc}\right)^{-2}\rm~erg~s^{-1}~cm^{-2}, \nonumber
\end{eqnarray}
where $d$ is the distance to the protostar and we use the distance to the nearby star-forming regions, such as Taurus molecular cloud and Rho Ophiuchi molecular complex.
This value is an order of magnitude lower than the sensitivity of CTA ($\sim10^{-13}\rm~erg~s~cm^{-2}$ for a 50-h integration: \href{https://www.cta-observatory.org/science/ctao-performance/#1472563157332-1ef9e83d-426c}{CTA website}). Thus, with these parameters, we cannot expect the VHE gamma-ray detection.
We numerically calculate the gamma-ray spectra from the protostellar flare with the method given in \citet{kab06} with $pp$ cross-section given in \citet{2014PhRvD..90l3014K}, and confirm this conclusion as shown in Figure \ref{fig:allband}. The result for our reference model (Model A) is indicated by the thick-red-solid line, which is much lower than the sensitivity curves of CTA (thin-magenta curves), respectively.

However, the gamma-ray luminosity strongly depends on the values of $B_{\rm rec}$ and $l_{\rm loop}$, which are largely unknown. 
 Figure \ref{fig:allband} also exhibits the GeV-TeV gamma-ray spectra from protostellar flares for other parameter sets tabulated in Table \ref{tab:results}.
 For a moderate case (Model B; high $B_{\rm rec}$ and $l_{\rm loop}$), the VHE flux  reaches the sensitivity of CTA.
 Since the duration of the gamma-ray emission is estimated to be $t_\gamma\approx t_{\rm rec}\sim10^3$ sec as tabulated in Table \ref{tab:results}, we can expect detection of VHE gamma rays if we stack $\sim180$ protostellar flares. Since the field-of-view (FoV) of CTA is $\sim3{\rm~deg}\times3\rm~deg$, they can monitor $\sim100-300$ YSOs simultaneously. Since 200-ks observation by Chandra (FoV$=0.29{\rm~deg}\times0.29\rm~deg$) found 71 YSO flares \citep{2001ApJ...557..747I,2003PASJ...55..653I}, we can expect several hundreds or several thousands of YSO flares within the FoV of CTA  during the 50-hour integration. Therefore, CTA could detect gamma rays if we perform coordinated observation with a wide-field X-ray satellite and a dedicated data analysis.

The protostellar flares may be able to accelerate CR protons more efficiently than solar flares. 
The amount of supra-thermal particles in the pre-shock region (i.e., reconnection outflow) is a key to determine the production efficiency of relativistic CRs, $\epsilon_p$. We infer that the protostellar flares will have a higher value of $\epsilon_p$ than solar flares, considering the geometry of the reconnecting magnetic fields. Theoretical investigations have suggested that magnetic reconnection with some amount of the guide field accelerates non-thermal particles less efficiently than reconnection without it \citep[e.g.][]{2021PhRvL.126m5101A}. Regarding the solar flares, it is expected that coronal reconnecting fields have a strong guide field before the flare onset because the coronal field is generally sheared by photospheric plasma motions \citep[e.g.,][]{2007ApJ...655L.117S,2017ApJ...850...39T}. The same will be true for the stellar flares, as their properties are similar to the solar ones \citep{2021ApJ...920..154G}. On the other hand, we consider protostellar flares powered by accretion \citep{2019ApJ...878L..10T}. In this framework, the protostar gains magnetic energy by getting the large-scale poloidal fields from the disk via accretion. 
In this case, the magnetic reconnection in the protostellar flares occurs with little guide fields (see Panel (2) of Figure \ref{fig:schematic})\footnote{Although a shear motion exists in the disk, the magnetic reconnection proceeds with a weak guide field component as long as the symmetry across the equatorial plane is not strongly violated \citep{2022ApJ...924L..32R}.}. The reconnection with such a configuration will more efficiently produce supra-thermal particles in the reconnection outflow \citep{2021PhRvL.126m5101A,2021PhRvL.127r5101Z}. Such high-energy particles will be easily re-accelerated via the diffusive shock acceleration \citep{2018JPlPh..84c7101C}. Therefore, protostellar flares can have a higher value of $\epsilon_p$ than solar flares.
Note that protostellar flares in the magnetospheric accretion paradigm are powered by a different mechanism \citep[e.g.][]{1996ApJ...468L..37H,2004Ap&SS.292..573U}, where the main energy source is twisting of the stellar field by the rotating disk.

Considering the above argument, we also plot the hadronic gamma-ray spectrum for an optimistic case with $\epsilon_p=0.1$ (Model C) in Figure \ref{fig:allband}. In this case, CTA can detect gamma rays from a single protostellar flare without the help of an X-ray satellite. We expect the enhancement of CR production, compared to solar flares, only for the accretion-driven flare scenario. Thus, future VHE gamma-ray observations of protostars may be useful to probe the triggering mechanism of protostellar flares and non-thermal phenomena occurring there.

Hadronic gamma-ray emission is inevitably accompanied by high-energy neutrinos, whose energy and differential luminosity and energy can approximately be written by $E_\nu\sim(1/2)E_\gamma$ and $E_\nu L_{E_\nu}\approx (1/2)E_\gamma L_{E_\gamma}$, respectively. We evaluate the neutrino detectability by current and near-future km$^3$ detectors \citep{KM3NeT16a,2014NIMPA.742...82A,2020NatAs...4..913A}, but the neutrino fluence from the protostellar flares are too low to detect. We need to stack $\sim10^5$ protostellar flares to detect a single neutrino with km$^3$ detectors, and hence, we cannot expect any neutrino detection even with the future 10-km$^3$ detectors \citep{Aartsen:2020fgd,2022arXiv220704519Y}.

\section{Leptonic emission}\label{sec:leptonic}

In this section, we discuss the detectability of emissions by relativistic electrons.
At the termination shock, primary CR electrons can be accelerated together with protons.
The relativistic CR electron production by the shock is likely more inefficient than protons, and PIC simulations suggest that the number density at a given energy can be as low as $K_{ep}=N_{E_e}/N_{E_p}\sim0.001-0.01$~\citep{PCS15a}, where $N_{E_e}$ is the number spectrum of the relativistic electrons and $E_e$ is the electron energy.
Then, the CR electron luminosity at the GeV energy is estimated to be
\begin{eqnarray}
& & L_e\approx\frac{\epsilon_e}{\epsilon_p}L_p\simeq 1.7\times10^{25}\label{eq:Le}\\
 &\times&\epsilon_{e,-4.5}l_{\rm loop,9.5}^2B_{\rm rec,2}^3n_{\rm rec,9}^{-1/2}\eta_{\rm rec,-1.5}f_{\rm nl,0.5}\rm~erg~s^{-1} \nonumber,
\end{eqnarray}
where $\epsilon_e$ is the efficiency of CR electron production. 
We use $\epsilon_e = 10^{-1.5}\epsilon_p$ as a fiducial value \citep[e.g.,][]{2012ApJ...758...81M}, but the value of $\epsilon_e$ in non-relativistic shocks is highly uncertain.
These relativistic electrons emit broadband photons via bremsstrahlung, synchrotron, and inverse Compton emissions.

\subsection{Solar flares}

The existence of non-thermal electrons is confirmed in solar flares \citep[e.g.,][]{2011SSRv..159..107H,2018SSRv..214...82O}, although the observed signals in hard X-ray bands are produced by non-relativistic electrons.
The relativistic electrons are confined in the flare loop with a timescale of $t_{\rm esc}$ (see Equation (\ref{eq:tesc})). We find that $t_{\rm esc}$ is much shorter than all the cooling timescales by bremsstrahlung, synchrotron, and inverse Compton processes.
Therefore, the maximum energy of electrons should be the same as that of protons: $E_{e,\rm max}=E_{p,\rm max}$. 
Although this value of $E_{e,\rm max}$ seems to be higher than expected, the current hard X-ray and soft gamma-ray observations cannot constrain the maximum energy of electrons, because the photon spectrum exhibits no cutoff features below the MeV range \citep[e.g.,][]{2003ApJ...595L..69L}. 

We find that the emissions from relativistic electrons in the flare loops are challenging to detect.  The bremsstrahlung and inverse Compton processes mainly produce GeV gamma rays, but they are fainter than the hadronic gamma rays. The synchrotron radiation spectrum has a peak at the soft X-ray band, which is easily overshone by thermal free-free emission by the evaporation plasma. 

Nonthermal electrons produced in the solar corona move to the chromosphere and lose their energies via bremsstrahlung and Coulomb interactions \citep{1971SoPh...18..489B,1978ApJ...224..241E}. 
This picture is consistent with the hard X-ray observations at footpoints of the flare loop \citep[e.g.,][]{2011SSRv..159..107H}. 
We should note that we cannot detect the bremsstrahlung photons by the relativistic electrons because of the relativistic beaming effect. Almost all photons are beamed toward the Sun due to their inefficient isotropization.

\subsection{Protostellar flares}

For protostellar flares, $B_{\rm rec}$ and $l_{\rm loop}$ are higher than those for solar flares. This leads to longer $t_{\rm esc}$ and $t_{\rm sh}$ and shorter $t_{\rm syn}$.
We find that the synchrotron cooling is efficient for protostellar flares and other emission components are negligible for electrons of $E_e\gtrsim5$ MeV. The maximum energy of CR electrons is given by the balance between synchrotron cooling and acceleration. The synchrotron cooling time is given as  $t_{\rm syn}=6\pi m_ec/(\sigma_TB_{\rm rec}^2\gamma_e)$, where $\gamma_e$ is the electron Lorentz factor and $\sigma_T$ is the Thomson crosssection. This leads to the maximum electron Lorentz factor and synchrotron cutoff energy of
\begin{equation}
 \gamma_{e,\rm max}=\sqrt{\frac{9\pi e\beta_A^2}{10\sigma_T\xi B_{\rm rec}}}\simeq 7.4\times10^4 B_{\rm rec,3}^{1/2}n_{\rm rec,10}^{-1/2}\xi^{-1/2}_{0.3},
\end{equation}
\begin{equation}
 E_{\gamma,\rm max}=\frac{h eB_{\rm rec}\gamma_{e,\rm max}^2}{2\pi m_ec}\simeq 63 B_{\rm rec,3}^{2}n_{\rm rec,10}^{-1}\xi^{-1}_{0.3}\rm~keV, 
\end{equation}
where $\beta_A=V_A/c$ and $h$ is the Planck constant.
Thus, non-thermal electrons emit broadband emission below hard X-ray bands. Equating $t_{\rm syn}$ and $t_{\rm esc}$, we obtain the cooling electron energy of $\gamma_{e,c}\simeq23 l_{\rm loop,11}^{-1}B_{\rm rec,3}^{-2}\eta_{\rm turb,1}^{-1}$. The electron number spectrum has a break at $\gamma_e=\gamma_{e,c}$: $N_{\gamma_e}\propto \gamma_e^{-s}$ for $\gamma_e<\gamma_{e,c}$ and $N_{\gamma_e}\propto \gamma_e^{-s-1}$ for $\gamma_e>\gamma_{e,c}$ \citep[e.g.,][]{SPN98a,2020ApJ...904..188K}.

The synchrotron-self absorption (SSA) is also efficient at low energies. This process shapes a low-energy cutoff in the non-thermal synchrotron spectra. We also estimate the optical depth for SSA \citep[e.g.,][]{rl79}, 
\begin{eqnarray}
\tau_{\rm ssa}&=& l_{\rm loop}\alpha_{\rm ssa}\nu^{-(p_e+4)/2} \\
\alpha_{\rm ssa}&=&\frac{\sqrt3 e^3}{8\pi m_e}\left(\frac{3e}{2\pi m_e^3 c^5}\right)^{p_e/2}\left(\frac{\pi B_{\rm rec}}{4}\right)^{(p_e+2)/2}\nonumber\\
& &\times C_{\rm ssa} \Gamma\left(\frac{3p_e+2}{12}\right)\Gamma\left(\frac{3p_e+22}{12}\right),
\end{eqnarray}
where $C_{\rm ssa}=L_et_{\rm esc}E_{\rm min}^{p_e-2}/(l_{\rm loop}^2l_{\rm nl})$ is the normalization factor for non-thermal electron spectrum, $p_e\approx s_{\rm inj}+1$ is the electron spectral index after the cooling break, $\Gamma(x)$ is the Gamma function.
Setting $\tau_{\rm ssa}=1$, we estimate the SSA cutoff frequency to be
\begin{eqnarray}
 \nu_{\rm ssa} &=& \left(\alpha_{\rm ssa}l_{\rm out}\right)^{2/(p_e+4)}\\
 &\simeq& 54 B_{\rm rec,3}^{11/7}n_{\rm rec,10}^{-1/7}l_{\rm loop,11}^{2/7}\eta_{\rm rec,-1.5}^{2/7}\eta_{\rm turb,1}^{2/7}\epsilon_{e,-4.5}^{2/7}  {\rm~GHz}, \nonumber
\end{eqnarray}
where we use $p_e=3$ for the estimate.
We find that $\nu_{\rm ssa}\sim1$ GHz for X-class solar flares and $\nu_{\rm ssa}\sim100$ GHz for typical protostellar flares, as tabulated in Table \ref{tab:results}.
For the range of our investigation, the SSA cutoff frequency is close to or higher than the cooling frequency, $\nu_c=E_{\gamma,c}/h$. Then, the synchrotron spectrum can be approximately written by
\begin{eqnarray}
 E_\gamma L_{E_\gamma}\approx L_{\rm syn}\left\{
\begin{array}{ll}
 \left(\frac{E_\gamma}{E_{\gamma,\rm min}}\right)^{2}\left(\frac{E_{\gamma,\rm min}}{E_{\gamma,\rm ssa}}\right)^{5/2} & (E_\gamma<E_{\gamma,\rm min}) \\
 \left(\frac{E_\gamma}{E_{\gamma,\rm ssa}}\right)^{5/2} & (E_{\gamma,\rm min}<E_\gamma<E_{\gamma,\rm ssa}) \\
 1 & (E_{\gamma,\rm ssa}<E_\gamma<E_{\gamma,\rm max}) \\
\end{array}
\right.
\end{eqnarray}
\begin{eqnarray}
 L_{\rm syn}\approx f_{\rm bol}L_e\simeq 1.7\times10^{29}l_{\rm loop,11}^2B_{\rm rec,3}^3n_{\rm rec,10}^{-1/2}\\
 \times\epsilon_{e,-4.5}\eta_{\rm rec,-1.5}f_{\rm nl}f_{\rm bol,-1}\rm~erg~s^{-1} \nonumber
\end{eqnarray}
where $E_{\gamma,\rm syn,min}=h eB/(2\pi m_ec)$, $E_{\gamma,\rm ssa}=h\nu_{\rm ssa}$, and $f_{\rm bol}=1/\ln(E_{\gamma,\rm max}/E_{\gamma,\rm ssa})$ is the bolometric correction factor \citep[e.g.,][]{2022arXiv220206480K}.

For a quantitative prediction, we numerically integrate Equation (\ref{eq:transport}) with a steady state assumption to obtain the non-thermal electron spectrum, and calculate the synchrotron spectrum by the method in \citet{2020ApJ...904..188K,2020ApJ...905..178K}. We take account of the SSA process by  $E_\gamma L_{E_\gamma}|_{\rm obs}=f_{\rm atn}E_\gamma L_{E_\gamma}|_{\rm thin}$, where $E_\gamma L_{E_\gamma}|_{\rm thin}$ is the synchrotron spectra without the SSA process and $f_{\rm ssa}\approx (1-\exp(-\tau_{\rm ssa}))/\tau_{\rm ssa}$ is the absorption factor. Here, we ignore the heating by the SSA process for simplicity, which may modify the electron distribution and synchrotron spectrum around the SSA frequency \citep{1988ApJ...334L...5G,1991MNRAS.252..313G}.

Figure \ref{fig:allband} shows the resulting radio to UV spectra from protostellar flares. The SSA cutoff appears around $100-1000$ GHz in our models. As seen in the figure, the synchrotron spectra can be observable above the foreground signals (disk and jet components) for all the models in 10--100 GHz bands, despite that the SSA is effective in the band.  T-Tauri stars (class II/III YSOs) exhibit flares in mm and sub-mm bands \citep{2003ApJ...598.1140B,2006A&A...453..959M,2019ApJ...871...72M, 2021ApJ...915...14N}. These mm and sub-mm flares can be interpreted as the synchrotron emission by relativistic electrons accelerated during the stellar flares. The observed luminosities for mm/sub-mm flares are $\nu L_\nu\sim10^{29}-10^{31}\rm~erg~s^{-1}$, which is consistent with our model predictions.
The spectral break by synchrotron emission in mm/sub-mm bands can be useful to estimate the magnetic field strength and/or non-thermal electron number density, as shown in Figure \ref{fig:allband}. This method is used in solar flares \citep{2022Natur.606..674F}. Future multi-band surveys together with detailed modeling of non-thermal emission will be able to unravel the non-thermal phenomena during the YSO flares.
We should note that the triggering mechanism of T-Tauri flares is expected to be similar to that of solar flares, because T-Tauri flares have X-ray features similar to those of main sequence stars \citep{2021ApJ...916...32G}. However, T-Tauri stars likely have strong magnetic fields and large flare loops, which are comparable to those for protostellar flares. Therefore, the mm/sub-mm emission from T-Tauri flares should be similar to those for our Model A and B.

The timescales of radio flares observed in mm/sub-mm bands vary greatly among the events \citep[e.g.,][]{2021ApJ...915...14N}. Some show minutes-scale variability, which is consistent with our model prediction. 
Recently, minutes-scale stellar flares are also observed in optical band \citep{2022PASJ...74.1069A}, although these are from M dwarfs. 
Others show variability on a timescale of a few days, which is probably caused by time-variable accretion.

The synchrotron fluxes in the infrared to optical bands are much lower than that by the protostar. The synchrotron spectrum is stronger than the protostar component in UV and very soft X-ray ranges, but these signals are completely attenuated due to dense envelopes. The synchrotron spectrum extends to hard X-ray and MeV ranges as seen in Figure \ref{fig:allband}. However, the hard X-ray flux is lower than the thermal X-ray emission (see Section \ref{sec:protostar}), and the MeV flux is lower than the sensitivity of future satellites, such as e-ASTROGAM \citep{2017ExA....44...25D}, AMEGO \citep{2019BAAS...51g.245M}, and GRAMS \citep{2020APh...114..107A}.

\section{Discussion}\label{sec:discussion}

\subsection{Cosmic-rays in molecular clouds}

\citet{2020PhRvD.101h3018A,2020ApJ...901L...4B} found that the CR density above $\sim10$ GeV in some molecular clouds is higher than that in the local interstellar medium (ISM) by a factor of a few. This may indicate the existence of sources of CRs of $E_p\gtrsim100$ GeV in molecular clouds. On the other hand, HAWC Collaboration also put constraints on multi-TeV gamma-ray flux from giant molecular clouds \citep{2021ApJ...914..106A}, implying that there are no strong sources of CRs of $E_p\gtrsim10$ TeV. Our model predicts that protostellar flares can accelerate CRs above 10 TeV with a total CR energy of $\mathcal{E}_p\sim10^{35}-10^{38}$ erg, depending on the model parameters of protostellar flares. If some fraction of CRs escape from the flare loop, they diffuse into the molecular clouds and emit GeV-TeV gamma rays there via hadronuclear interactions. Also, these CRs may be ionization and heating sources of molecular clouds \citep{UN81a,2009A&A...501..619P,2021ApJ...915...43F}. Indeed, estimates of the CR ionization rate by interstellar chemistry modeling indicate that CR energy densities in dense clouds should be higher than that in local ISM \citep{2020SSRv..216...29P}.  Below, we roughly estimate the CR energy density in a molecular cloud provided by protostellar flares.

 We consider that a typical molecular cloud of size $R_{\rm cl}\sim10$ pc contains $N_{\rm ps}\sim100$ protostars, as in L1688 in Rho Ophiuchi \citep{2008hsf2.book..351W}. 
The flare interval timescale on a single protostar will depend on the details of the interaction between the protostar and the innermost disk. 
The Chandra Orion Ultradeep Project (COUP), which was a two-week X-ray monitoring campaign by Chandra toward the Orion Nebula Cluster, found that a high detection rate ($\gtrsim 50$\%) of X-rays in class-I protostars \citep{2008ApJ...677..401P}. This suggests that protostars produce at least one huge flare per two weeks.
We estimate the differential CR generation rate in the molecular cloud to be $E_p L_{E_p}\approx \mathcal{E}_pN_{\rm ps}f_{\rm bol}f_{\rm esc}/t_{\rm intvl}\sim 10^{30}-10^{33}f_{\rm esc}f_{\rm bol,-1} \rm~erg~s^{-1}$, where $f_{\rm esc}$ is the escape fraction from the flare loop, $f_{\rm bol}\approx 1/\ln(E_{p,\rm max}/\rm GeV)$ is the bolometric correction factor, and we approximate  $s\approx2$. Balancing the CR injection to and escape from the molecular cloud, the differential CR energy density is estimated to be \citep[e.g.,][]{KMM18a}
\begin{eqnarray}
 E_pU_{E_p} &\sim& \frac{E_pL_{E_p}}{R_{\rm cl}D_p}\\
&\sim&7\times10^{-8}-7\times10^{-5}\left(\frac{E_p}{\rm TeV}\right)^{-1/3}\rm eV~cm^{-3}\nonumber
\end{eqnarray}
where we use a canonical diffusion coefficient of $D_p\sim 3\times10^{29} (E_p/{\rm TeV})^{1/3}\rm~cm^2~s^{-1}$ \citep[e.g.,][]{SMP07a}. These values are much lower than the CR energy density in ISM, $U_{\rm CR,ISM}\sim 3\times10^{-3}\rm~eV~cm^{-3}$ at 1 TeV \citep[e.g.,][]{2015PhRvL.114q1103A}. 
Thus, protostellar flares are unlikely to be the source of the CR excess in molecular clouds, unless the diffusion coefficient in molecular clouds is orders of magnitude lower than the canonical value.

\subsection{Cosmic rays in protoplanetary disks}

Cosmic rays accelerated at the flare can be the ionization source in protoplanetary disks. Previous studies assumed that protostellar flares produce CRs of MeV-GeV energies \citep[e.g.,][]{2002ApJ...572..335F,2018A&A...614A.111P}. These CRs can be an important ionization source of the disk surface \citep{tur09,rab17,fra18}, or even at the midplane if we assume efficient diffusion \citep{rod20}. Our model suggests that  protostellar flares produce CRs of much higher energies ($\sim0.1-1$ TeV). These CRs have a higher penetration power and diffuse faster than the lower-energy CRs. These features are advantageous for ionizing the disk midplane. However, \cite{2022ApJ...937L..37F} recently showed that intrusion of CRs to the disk midplane is more challenging than previously thought because CRs usually propagate along the sheared magnetic field configuration \citep[see also][for discussions on the propagation of CRs into a disk]{2018ApJ...863..188S}. The ionization rate by the CRs from protostellar flares should be investigated carefully by considering the propagation of high-energy CRs in a realistic magnetic field geometry.

CRs from protostars can produce short-lived radioactive nuclei, e.g., Al$^{26}$ and Be$^{10}$, which can be an important ionization source at the inner part of protoplanetary disks. Previous studies considered CR production at accretion shocks \citep{2020ApJ...898...79G} or utilized the empirical relation obtained from solar flare observations \citep{2019A&A...624A.131J}. Our model proposes that accretion-driven protostellar flares can also be an efficient CR accelerator, and the effects of these CRs should be examined near future.

\subsection{Attenuation of TeV gamma rays}

VHE gamma rays can be attenuated by the two-photon interactions ($\gamma\gamma\rightarrow e^+e^-$) if copious low-energy photons exist. We estimate the optical depth for two-photon interactions, $\tau_{\gamma\gamma}$. The flare loop is surrounded by the protostar and optically thick accretion disk, both of which emit thermal photons. Assuming that the temperatures of the protostar and the disk are $T_*=4000$ K, the typical energy of the thermal photons is $E_{\gamma,*}\approx 2.8 k_BT_*\sim1 (T_*/4000\rm~K)$ eV. Then, gamma rays of $E_\gamma\sim1(T_*/4000\rm~K)^{-1}$ TeV efficiently interact with the thermal photons. The number density of the thermal photons is estimated to be $n_{\gamma,*}\approx a_{\rm rad}T_*^4/E_{\gamma,*}\simeq1.3\times10^{12}(T_*/4000\rm~K)^3~cm^{-3}$. This results in $\tau_{\gamma\gamma}\approx 0.2\sigma_T n_{\gamma,*}l_{\rm loop}\simeq0.017(T_*/4000{\rm~K})^3l_{\rm loop,11}$. Therefore, TeV gamma rays likely escape from the system without attenuation by two-photon interactions. The inner part of protoplanetary disks may be hotter than the protostellar surface, and it can reach $\sim2\times10^4$ K \citep{1993ApJ...415L.127P}. In this case, $\tau_{\gamma\gamma}>1$ is satisfied above $\sim0.2$ TeV, and detection of very-high-energy gamma rays from protostellar flares is challenging (see also \citealt{2011ApJ...738..115D} for discussion on $\tau_{\gamma\gamma}$ with a different photon field). 

\subsection{Implications for the driving mechanism of protostellar flares}

\citet{2009ApJ...691..823H} suggested that in the early evolutionary stage, a protostar does not have a surface convective layer depending on the accretion history. As the surface convective layer would be crucial for the magnetic activity mediated by the stellar dynamo, we cannot expect dynamo-origin flares in the absence of the convective layer. Thus, class-0 objects are unlikely to produce powerful flares by magnetospheric models. 
On the other hand, class-0 objects should exhibit strong flaring activities in our accretion-driven flare model. Since class-0 objects are highly obscured in soft X-rays \citep{2020A&A...638L...4G}, we need to rely on non-thermal signatures in mm/sub-mm and gamma-ray bands to probe the flaring activities in class-0 objects. A monitoring campaign of nearby class-0 objects would test the stellar evolution scenario and unravel the mechanism of flaring activities of protostars.

Our scenario of protostellar flares creates a cavity in the inner part of the protoplanetary disk. This may induce time variability in the near-infrared band, with a typical timescale of a few days \citep{2019ApJ...878L..10T}.
Such a time variability is observed in infrared by Hershel \citep{2012ApJ...753L..35B}.
The dust in protoplanetary disks can be observed in GHz-radio bands, and the angular resolution of ngVLA can be $0.01-0.1$ AU scale for a distance of $\sim150$ pc. Hence, ngVLA may directly resolve the cavity. In our scenario, the size of the cavity may be larger than the magnetospheric models, and thus, future radio observations coordinated with X-ray satellites will reveal the protostellar flare mechanism.

\section{Summary}\label{sec:summary}

We developed models for the gamma-ray emissions from solar and protostellar flares.
We developed the gamma-ray emission model for solar flares based on a standard scenario driven by magnetic reconnection in the solar corona,
and the model for protostellar flares assumes the flare mechanism proposed by \citet{2019ApJ...878L..10T}. In both models, we expect the formation of X-ray luminous flare loops on the stellar surface as a result of magnetic reconnection. We hypothesize that CR protons are produced at the termination shock and are injected to the flare loops. We examine the high-energy gamma-ray production via hadronuclear interactions with the evaporation plasma. Our model can reproduce the gamma-ray data of solar flares and roughly be consistent with the X-ray data of the same flare.
We applied this model to protostellar flares. As a result, we found that CTA can detect VHE gamma rays from a single protostellar flare if the CR production is more efficient in protostellar flares than solar flares (Model C). If the CR production efficiency is the same as that for solar flares (Models A and B),  VHE gamma-ray detection is challenging near future. 
Emission by non-thermal electrons accelerated together with protons are also detectable in mm/sub-mm bands for all the models. The duration of the mm/sub-mm flare is $t_\gamma\sim10^3$ sec, while the time interval between the flare could be  $t_{\rm intvl}\gtrsim10^6$ sec (see Section \ref{sec:discussion} for the time interval). The duty cycle of the mm/sub-mm flare is $t_\gamma/t_{\rm intvl}\lesssim10^{-3}$. Thus, dedicated strategic observations are necessary to identify the mm/sub-mm flares from protostars.


\acknowledgements
We thank Kazumi Kashiyama for fruitful discussion. This work is partly supported by KAKENHI No. 22K14028 (S.S.K.), 21H04487 (S.S.K., S.T. and K.T.), 22K14074, and 22H00134 (S.T.). S.S.K. acknowledges the support by the Tohoku Initiative for Fostering Global Researchers for Interdisciplinary Sciences (TI-FRIS) of MEXT's Strategic Professional Development Program for Young Researchers.

\bibliographystyle{aasjournal}
\bibliography{ssk}

\listofchanges

\end{document}